\documentstyle[twocolumn,aps,epsf]{revtex}

\newcommand{\ewxy}[2]{\setlength{\epsfxsize}{#2}\epsfbox[10 60 640 570]{#1}}

\begin{document}

\draft        

\title{The Effectiveness of Non-Perturbative O($a$) Improvement in Lattice QCD}
\author{R. G. Edwards, U. M. Heller and T. R. Klassen}
\address{SCRI, Florida State University, Tallahassee, FL 32306-4130}


\maketitle

\begin{abstract}
The ALPHA collaboration has determined the O($a$) improved
Wilson quark action for lattice spacings $a\leq 0.1$~fm, in the quenched
approximation.     We extend this result to coarser lattices, 
$a\leq 0.17$~fm, and calculate the hadron spectrum on them. 
The large range of lattice spacings obtained by combining our results
with earlier ones on finer lattices,  allow  us to present 
  a convincing demonstration
of the efficiency of non-perturbative O($a$) improvement.
We find that scaling violations of 
the hadron masses studied drop from $30-40\%$ 
for the
unimproved  Wilson action on the coarsest lattice to only $2-3\%$.

\end{abstract}

\pacs{PACS: 12.38.Gc, 11.15.Ha}  

\narrowtext

{\bf Introduction.}
To measure standard model parameters, like CKM matrix elements and quark 
masses, and to find signatures of new physics, 
accurate knowledge of weak matrix elements between 
hadronic states is required. Lattice QCD
is the only systematically improvable method of obtaining this information.
The high cost of lattice QCD simulations has lead to a renewed appreciation
of the fact that progress in this field 
depends to a large extent on the successful 
use of  ``improvement'' ideas
(see the proceedings of the last few Lattice Field Theory conferences,
e.g.~\cite{LAT97} for the last one). The reason is the following.
To avoid doublers, 
the Wilson-type quark actions most commonly used in simulations
must break chiral symmetry at some level.
On the quantum level at least, this violation will generically 
occur at leading order in the lattice spacing, O($a$).
These errors therefore decrease only slowly with the
lattice spacing and their absolute value is large, as experience has
shown. To perform accurate and reliable continuum extrapolations would
require the use of very fine lattices, for which simulations are very
expensive. 

A much better approach\cite{Sym} is to 
correct the discretization  errors of a lattice action
by adding higher-dimensional (irrelevant) operators to the action 
which reproduce the effects of the UV modes omitted on the lattice.
Trying to do so  perturbatively 
did initially not appear to be a significant improvement.
It was then realized~\cite{LM} that
large perturbative corrections arise due to lattice-specific
``tadpole'' graphs, and can be corrected by a mean-field type method.
Nevertheless, 
as a resummation of certain graphs in perturbation theory, 
this approach can basically
only reduce quark errors from O($a$) to order $g^2 a$ or  
$g^4 a$. This is only a logarithmic suppression compared to
O($a$) and would still require the inclusion of at least  
$g^4 a$ (say)
and $a^2$ terms in an honest continuum extrapolation of the 
discretization errors. This leads to large errors and potentially
unstable fits.

To eliminate the O($a$) errors of spectral quantities there is
only one term that has to be added to the Wilson QCD action~\cite{SW}.
The gauge action retains the standard plaquette form, and the 
quark action (density) becomes
\begin{equation}\label{Action}
\bar{\psi}(x) \left[ \sum_\mu (\gamma_\mu \nabla_\mu -
                        {\textstyle {1\over 2}} a\Delta_\mu)
 \, - \, {\textstyle {1\over 4}} a \omega 
      \sum_{\mu\nu} \sigma_{\mu\nu} F_{\mu\nu}\right] \psi(x) .
\end{equation}
Here $\nabla_\mu$ and $\Delta_\mu$ are the standard covariant first,
respectively, second order lattice derivatives. The new 
$\sigma\!\cdot\! F$ term involves the $\sigma$-matrices
$\sigma_{\mu\nu}=-{i\over 2} [ \gamma_\mu,\gamma_\nu ]$ and a 
discretization of the field strength $F_{\mu\nu}$. Inspired by the 
form of its most popular discretization, this term is also known as the
``clover'' term, and the coefficient $\omega$ as the clover
coefficient. To eliminate O($a$) errors, $\omega$ has to be
determined as a function of the gauge coupling $g$.

A great step forward was recently taken by the ALPHA 
collaboration~\cite{ALPHA}, which used the chiral Ward identity
as an improvement condition to determine the non-perturbative
value of $\omega$.
This was accomplished in the context of the Schr\"odinger 
functional~\cite{LNWW},
where one imposes fixed boundary conditions      on the gauge
and fermion fields       in the time
direction, and can then work at zero, or at least small,
quark masses.
The ALPHA collaboration determined improvement coefficients for
lattice spacings of about $a\leq 0.1$~fm (more precisely,
$\beta\!\equiv\! 6/g^2 \geq 6.0$ in standard notation).

Since one needs a minimum of three or four
reasonably separated lattice spacings to perform accurate and
reliable continuum extrapolations, this goal will not easily be 
accomplished, even in the ``quenched'' approximation 
(where quark loops are ignored, and to which the above results
refer), if only lattices of spacing $0.1$~fm and less are considered.
We will explicitly see this below.
We have therefore attempted to extend the results of the ALPHA
collaboration to coarser lattices.

{\bf Chiral Symmetry Restoration at O($a$).}
Consider QCD with (at least) two flavors of mass-degenerate quarks.
The idea~\cite{ALPHA}
for determining the clover coefficient is that chiral symmetry
will hold only if its Ward identity is satisfied
as a {\it local operator equation}. In Euclidean space this means that
the PCAC relation between the iso-vector axial
current and the pseudo-scalar density,
\begin{equation}\label{PCAC}
  \langle \partial_\mu A_\mu^b(x) \, {\cal O} \rangle = 
     2m \, \langle P^b(x) \, {\cal O} \rangle
\end{equation}
should hold for all operators ${\cal O}$, 
boundary conditions, $x$ (as long as $x$ is not in the support of ${\cal O}$),
and also for volumes that are not necessarily large in physical units.
 More precisely, it should hold {\it with the same mass} $m$ 
up to $a^2$ errors.
 This will only be the case for the correct value of 
the clover coefficient. 

Several issues have to be addressed before this idea can be implemented
in practice. First of all, even though here we can ignore the multiplicative 
renormalization of $A_\mu^b$ and $P^b$,    
there is an additive correction to $A_\mu^b$ at O($a$),
\begin{eqnarray}\label{AP}
   P^b(x)     &\, \propto\, & 
                \bar{\psi}(x)         \gamma_5 
                      {\textstyle {1\over 2}} \tau^b \psi(x) \: , \nonumber \\
   A^b_\mu(x) &\, \propto\, & 
                 \bar{\psi}(x) \gamma_\mu \gamma_5 
                          {\textstyle {1\over 2}} \tau^b \psi(x)
                     + a \, {c_A} \, \partial_\mu P^b(x)         \:.
\end{eqnarray}
The determination of $\omega$ is therefore tied in with that of the
axial current improvement coefficient $c_A$. 
Since in principle~(\ref{PCAC}) provides infinitely many conditions,
this is not a fundamental difficulty. How to solve it in practice 
is discussed in~\cite{ALPHA,TKSF}.

Note that $\omega$ and $c_A$ have an O($a$) ambiguity;
different improvement conditions will give somewhat
different values for $\omega$ and $c_A$. Instead of assigning a
systematic error to
$\omega$ and $c_A$ one should choose a specific, ``reasonable'' improvement
condition ---  the difference 
in observables from this versus some other choice   is
guaranteed to extrapolate away like O($a^2$) in the continuum limit.

For various 
reasons it is preferrable to impose the PCAC
relation at zero quark mass.
Due to zero modes this is not possible with periodic boundary conditions; 
the quark propagator would diverge.  Another reason to abandon periodic 
 boundary conditions is that to be sensitive to
the value of $\omega$ it would be highly advantageous to have a background
field present; it couples directly to the clover term. 

The Schr\"odinger functional
provides a natural setting to implement these goals.
By choosing suitable boundary conditions at the ``top'' ($x_0=T$)
and ``bottom'' ($x_0=0$) of the lattice world, one induces a chromo-electric
classical background field, and, at least at weak coupling, the quark operator
has no zero modes at vanishing quark mass
(the lowest eigenvalue being of order $1/T$).

We must now choose a specific improvement condition for $\omega$.
The idea is that by averaging Eq.~(\ref{PCAC}) over spatial volume,
each choice of ${\cal O}$ defines an estimate $m_{{\cal O}}(x_0)$ of 
the current quark mass. Requiring the {\it difference}
$\Delta m(x_0) \equiv m_{{\cal O}_1}(x_0) - m_{{\cal O}_2}(x_0)$
 for two specific ${\cal O}_1$ and
${\cal O}_2$ to vanish for suitable $x_0$, provides a non-perturbative 
condition to fix $\omega$. 
In practice, one calculates all required correlation functions in
a Monte Carlo simulation for several trial values of $\omega$ and finds
the zero crossing of $\Delta m(x_0)$ (more precisely, one should equate
it to its small, order $a^2$ tree-level value). This determines the
non-perturbative $\omega$, with some statistical error, 
for the chosen value of the gauge coupling.

A natural choice of ${\cal O}_1$ and ${\cal O}_2$ is provided by 
{\it boundary fields}~\cite{ALPHA}
associated to the lower and upper boundaries of the lattice. We will not
elaborate on these and other choices one makes in the calculation of
$\omega$; the details have been discussed in the 
literature~\cite{ALPHA,TKSF} and the specifics of the simulations described
here can be found in~\cite{EHKlat97}. 

\begin{figure}
\ewxy{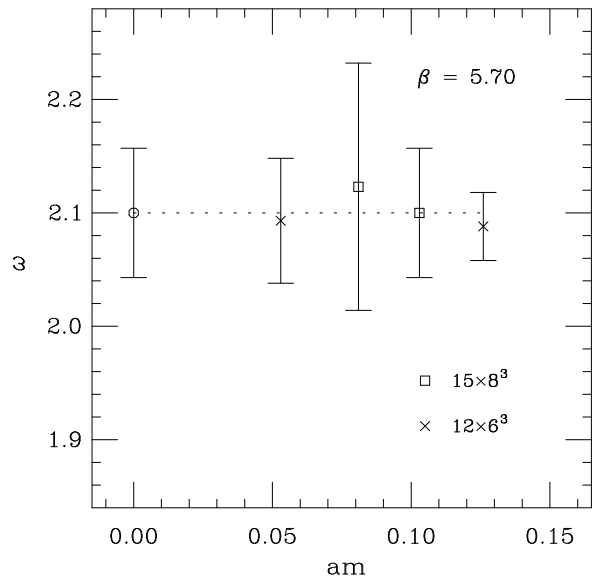}{100mm}
\caption{The non-perturbative clover coefficient as function of quark mass and 
         volume for $\beta=5.7$. We also show our choice of the $m=0$ value.}
\label{w_b570}
\end{figure}

\begin{figure}
\ewxy{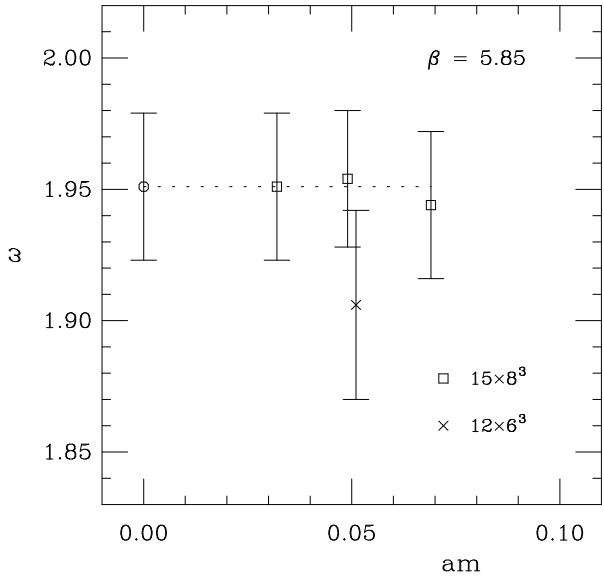}{100mm}
\caption{As in figure~\protect\ref{w_b570} for $\beta=5.85$.}
\label{w_b585}
\end{figure}

We have to mention, however, one important point. The above simulations
at different trial values of $\omega$ should be performed at a fixed
value of the quark mass (defined by, say, $m\equiv m_{{\cal O}_1}(z_0)$ 
for suitable $z_0$), preferrably zero. 
It turns out that in the quenched approximation
this is not possible on coarse lattices: Despite the non-periodic boundary
conditions one finds in practice that for roughly $\beta \leq 6.0$
one occasionally hits configurations, known as ``exceptional configurations'',
with an accidental \mbox{(near-)}zero mode, leading to a (near-)divergence of
the quark propagator.
(With periodic boundary conditions configurations with near-zero modes at
small quark mass exist for {\it any} finite $\beta$ in the 
quenched approximation; however, their frequency rapidly decreases at 
weak coupling.)   They can be avoided by using a larger quark mass, but
the question is to what extent this 
affects              the value of $\omega$. 
Fortunately, it turns out that the mass dependence of $\omega$ is 
extremely weak, so that one can reliably determine $\omega$ at larger masses.
This is illustrated in figures~\ref{w_b570} and~\ref{w_b585} 
for coarse lattices  (cf.~also \cite{EHKlat97}).

\begin{figure}
\ewxy{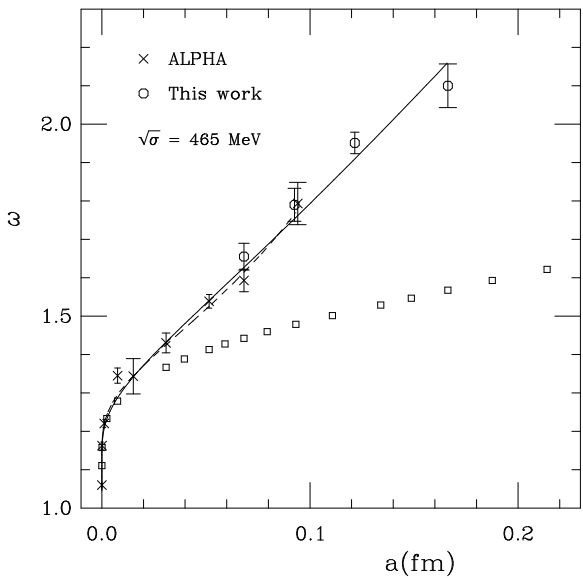}{100mm}
\caption{The measured non-perturbative clover coefficient and its
parameterization for $\beta\geq 5.7$ (solid line). The dashed line 
denotes the curve from~\protect\cite{ALPHA}. The tree-level tadpole
estimate from the plaquette is also shown ($\Box$). (Using the
mean-link in Landau gauge gives an estimate closer to the 
non-perturbative determination, cf.~\protect\cite{Tsukuba}.)  }
\label{w_Wil}
\end{figure}

For use of the non-perturbatively improved action in later simulations it
is advisable to present the results for $\omega$ in terms of a smooth
function of the gauge coupling. Combining the results of the ALPHA 
collaboration~\cite{Rainer} with our measurements for
$\beta\!=\! 5.7, 5.85, 6.0$ and $6.2$,  we find that they
can be represented by
\begin{equation}\label{wofg2}
\omega(g^2) = { 1 - 0.6084\,g^2 - 0.2015\,g^4 + 0.03075\, g^6\over
                1 - 0.8743\,g^2 }
\end{equation}
for $\beta\!\equiv\! 6/g^2 \!\geq\!5.7$.
This curve incorporates the one-loop perturbative result~\cite{w1loop}.
It never deviates by more
than 1.0\% from the curve presented in~\cite{ALPHA} for $\beta \geq 6.0$.
This is illustrated in figure~\ref{w_Wil}, where we used the 
parameterization of the string tension from~\cite{EHKsigma} 
to present the clover coefficient as a function of lattice spacing.

{\bf Hadron Spectrum.} 
To check how small scaling violations of spectral quantities are
after non-perturbative improvement of the action, we have calculated
the hadron spectrum using Eq.~(\ref{wofg2}) for 
$\beta\!=\! 5.7$ and $5.85$.
For a scaling check it is not necessary to consider
light hadrons. To avoid the uncertainties of the chiral extrapolation
we will instead consider hadrons at a pseudo-scalar to vector meson
mass ratio of $m_P/m_V=0.7$, corresponding
roughly to the strange quark.
This also avoids problems with exceptional configurations, which afflict
simulations at smaller masses on our coarsest lattice. We regard them as an 
essentially technical problem of Wilson-type quarks in the quenched approximation 
(it does not occur for full QCD or     staggered fermions), 
orthogonal to the issue of improvement.

Masses were obtained through two-exponential fits of correlators from
one under- and one over-smeared  source.
We used 400 configurations, statistically enhanced through the use of
sources constructed by superimposing different origins 
with random ${\relax{\sf Z\kern-.4em Z}}_3$  phases~\cite{Z3}.
Our results are given in table~\ref{table3}.   We also show
data from other groups on finer lattices, which we interpolated to
$m_P/m_V=0.7$. Since we can not do correlated fits of their  data,
we multiplied the naive error from interpolating fits with a factor
of 1.5. This gives values close to the actually measured errors 
for neighboring mass values. We hope that in the future it will become
customary to quote hadron masses interpolated to $m_P/m_V=0.7$ 
and perhaps 
a few other benchmark values (like 0.6 and 0.5). 
The results in~\cite{Gock,Mendes} were obtained using the parameterization
of $\omega$ from~\cite{ALPHA}, 
instead of Eq.~(\ref{wofg2}). We estimate that this changes
the masses by less than 0.4\%, which is negligible compared to the
current statistical errors.
The string tensions were taken from our interpolation 
formula~\cite{EHKsigma}, which is
based on recent precise measurements by us and others.
We assign~\cite{EHKsigma} these string tensions a 1\% (or smaller) error, 
that  can be added at the end.
We find that excellent fits to a $\,{\rm const}+a^2$
ansatz are possible, yielding $m_V/\sqrt{\sigma} = 2.351(20)$ and
$m_N/\sqrt{\sigma} = 3.466(36)$.

\begin{figure}
\ewxy{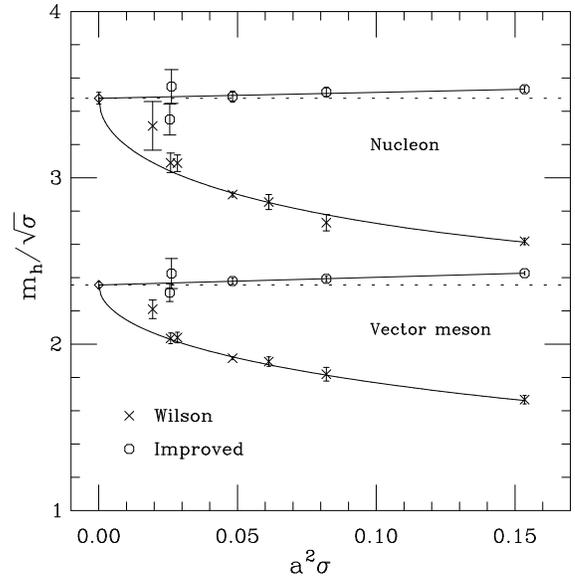}{100mm}
\caption{The hadron spectrum from Wilson and improved actions
at $m_P/m_V = 0.7$. Also shown are joint fits of both 
data sets (the first vector meson, respectively, nucleon
fit from table~\ref{fits}).}
\label{had_spec}
\end{figure}

We have also considered joint fits with
data for the standard quenched Wilson QCD action ($\omega=0$).
Results from different groups for seven couplings in the 
range $\beta\!=\!5.7 - 6.3$
have been conveniently collected in~\cite{Gock} (errors are treated 
similarly as above). In a joint fit we demand that the ans\"atze for
the improved and standard Wilson data intercept at the same point
in the continuum limit.  The results are shown in table~\ref{fits}
and figure~\ref{had_spec}. The joint fits agree perfectly with 
fits using only the improved action data.  For the Wilson
data it is necessary to have O($a$) and O($a^2$) terms in the ansatz to 
get a reasonable $Q$ in fits where $\beta=5.7$ is included. 
Fitting the Wilson data alone
yields fits that have either bad $Q$'s or large errors; they are
also not very stable under leaving out small (or large) $\beta$ 
points. This illustrates how difficult it is to perform reliable
continuum extrapolations with the Wilson action. Figure~\ref{had_spec}
also demonstrates that continuum extrapolations using only lattices with
$\beta \geq 6.0$ ($a^2\sigma < 0.05$ or about $a\leq 0.1\,$fm) 
would be quite expensive.

{\bf Conclusions.}
Figure~\ref{had_spec} is
impressive proof for the effectiveness of non-perturbative O($a$)
improvement: The scaling violations at $\beta=5.7$ are
reduced from 41\% to 3\% for the vector meson mass, and from 33\%
to 2\% for the ``nucleon'' mass. 
Even more important, the scaling in figure~\ref{had_spec} indicates
that O($a$) errors really have been eliminated from 
the improved action to high precision.  
 We should remark      that    without    the accurate string 
tension measurements from~\cite{EHKsigma} it would have been impossible
to reach this conclusion.

An analysis of the above data and some toy
examples shows that it is a factor of 100 or so cheaper to 
achieve a 1\% (say) error in the hadron masses using the improved instead
of the standard Wilson action.  Since
there is no fundamental difference in the improvement program
between quenched and full QCD, we      expect very large improvements also
in more realistic situations like
full QCD with lighter quark masses.


\vskip 2mm

\noindent {\bf Acknowledgments.}
This work is supported by DOE grants DE-FG05-85ER25000 and 
DE-FG05-96ER40979.
The computations in this work were performed on the workstation
cluster, the CM-2, and the new QCDSP supercomputer at SCRI. 

\vskip 2mm

\begin{table}
\caption{Simulation parameters, string tensions~\protect\cite{EHKsigma},
and results for the vector meson and ``nucleon'' (octet) masses
at $m_P/m_V=0.7$ for the non-perturbatively improved action.
\label{table3}
}
\begin{tabular}{llccll}
$\beta$  & $a\sqrt{\sigma}$ & Volume & $N_{{\rm confgs}}$ 
                            & $m_V/\sqrt{\sigma}$ & $m_N/\sqrt{\sigma}$\\ 
\tableline
5.7      & 0.3917 & $16^3\!\cdot\!32$ & 400 & 2.427(10)  & 3.532(17)\\
5.85     & 0.2863 & $16^3\!\cdot\!32$ & 400 & 2.392(16)  & 3.515(28)\\
6.0\tablenotemark[1]      
         & 0.2196 & $(16,\!24)^3\!\cdot\!32$  & $200\!-\!1000$ 
                                            & 2.380(17)  & 3.488(34)\\
6.2\tablenotemark[1]      
         & 0.1610 & $24^3\!\cdot\!48$ & 300 & 2.310(53)  & 3.351(93)\\
6.2\tablenotemark[2]      
         & 0.1610 & $24^3\!\cdot\!48$ & 104 & 2.425(91)  & 3.55(10)\\
\end{tabular}
\tablenotetext[1]{Ref.~\cite{Gock}}
\tablenotetext[2]{Ref.~\cite{Mendes}}
\end{table}

\vskip 2mm

\begin{table}
\caption{Fit parameters and confidence level $Q$ for joint and 
         separate fits of the improved and Wilson hadron mass data
         (at $m_P/m_V\!=\!0.7$) to ans\"atze of the form
         $m_V/\sqrt{\sigma} = V_0 + V_1\,a\sqrt{\sigma} + V_2\,a^2 \sigma$
         (for the vector meson; similarly for the nucleon).
\label{fits}
}
\begin{tabular}{l l c l l c l}
  & & \multicolumn{2}{c}{Improved} & \multicolumn{2}{c}{Wilson} & \\
$\beta_{{\rm min}}$  
         & ~~~$V_0$  & $V_1$  & ~~$V_2$  & ~~~$V_1$      & $V_2$  & $Q$ \\
\tableline
5.7      & 2.356(20) & 0      & 0.46(16) & $-2.2(2)$     & 1.1(5) & 0.29 \\
5.7      & 2.332(17) & 0      & 0.64(14) & $-1.82(8)$    & 0      & 0.09 \\
5.85     & 2.357(34) & 0      & 0.43(52) & $-2.0(2)$     & 0      & 0.26 \\
\tableline
5.7      & 2.351(20) & 0      & 0.50(16) &               &        & 0.74 \\
5.85     & 2.343(40) & 0      & 0.63(60) &               &        & 0.55 \\
\tableline
5.7      & 2.59(13)  &        &          & $-3.9(10)$    & 4.1(17)& 0.27 \\
5.7      & 2.286(32) &        &          & $-1.6(1)$     & 0      & 0.05 \\
5.85     & 2.392(65) &        &          & $-2.1(3)$     & 0      & 0.12 \\
\tableline
         & ~~~$N_0$  & $N_1$  & ~~$N_2$  & ~~~$N_1$      & $N_2$&        \\
\tableline
5.7      & 3.478(35) & 0      & 0.35(28) & $-3.1(3)$     & 2.2(6) & 0.38 \\
5.7      & 3.393(26) & 0      & 0.98(22) & $-2.1(1)$     & 0      & 0.01 \\
5.85     & 3.472(57) & 0      & 0.46(87) & $-2.6(3)$     & 0      & 0.56 \\
\tableline
5.7      & 3.466(36) & 0      & 0.44(28) &               &        & 0.52\\
5.85     & 3.425(69) & 0      & 1.1(10)  &               &        & 0.41\\
\tableline
5.7      & 3.97(22)  &        &          & $-6.7(16)$    & 8.2(27)& 0.87\\
5.7      & 3.312(38) &        &          & $-1.8(1)$     & 0      & 0.07\\
5.85     & 3.58(10)  &        &          & $-3.1(5)$     & 0      & 0.65\\
\end{tabular}
\end{table}

\end{document}